\documentclass[a4paper,11pt]{article}
\pdfoutput=1
\usepackage{jheppub}
\usepackage{amsmath}
\usepackage{amssymb}
\usepackage{amsfonts}
\usepackage{slashed}
\usepackage{array}
\usepackage{booktabs}

\newcommand{\cA}{\mathcal{A}}

\newcommand{\be}{\begin{equation}}
\newcommand{\ee}{\end{equation}}
\newcommand{\nn}{\nonumber}

\title{$\alpha'^2$-corrections to the bosonic string from an action-level double copy map}

\author[a,b]{Eric Lescano,}
\author[c,d]{Agustina Pereyra Grau,}
\author[d,e]{Jes\'us A. Rodr\'iguez}

\affiliation[a]{Jo\v{z}ef Stefan Institute, Jamova 39, 1000 Ljubljana, Slovenia}
\affiliation[b]{Instituto Superior T\'ecnico, Av.~Rovisco Pais 1, 1049-001 Lisboa, Portugal}
\affiliation[c]{Departamento de F\'isica, Universidad de Buenos Aires, Ciudad Universitaria, Pabell\'on 1, CABA, C1428ZAA, Argentina}
\affiliation[d]{Instituto de Astronom\'ia y F\'isica del Espacio (CONICET-UBA), Ciudad Universitaria, Pabell\'on IAFE, CABA, C1428ZAA, Argentina}
\affiliation[e]{Universidad Argentina de la Empresa (UADE), Lima 717, Buenos Aires, Argentina}

\emailAdd{eric.lescano@ijs.si}
\emailAdd{apereyragrau@iafe.uba.ar}
\emailAdd{jarodriguez@iafe.uba.ar}

\abstract{
The low-energy effective action of the bosonic string contains higher-derivative corrections of the form Riem$^2$ and Riem$^3$ to the Einstein--Hilbert action, together with the associated Kalb--Ramond couplings. We construct an action-level double copy description of these corrections up to cubic order in the fields. Two ingredients are required: a higher-derivative gauge-theory Lagrangian, which up to field redefinitions is controlled by a single $F^3$ coupling, and a deformation of the color-to-kinematics replacement of the structure constants by the corresponding three-momentum tensor. We show that no higher-derivative gauge-theory operator can produce the Riem$^3$ vertex with the standard replacement rule, and that the matching conditions for the gauge-theory coupling and the deformation of the map admit a unique solution, a double root, which renders the cubic action left--right symmetric, of the form $(\mathrm{YM}+F^3)\otimes(\mathrm{YM}+F^3)$. Without further input, the resulting map reproduces the cubic $\mathrm{Riem}\,H^2$ coupling of the bosonic string at order $\alpha'$, the absence of six-derivative cubic Kalb--Ramond couplings at order $\alpha'^2$, and the absence of interactions odd in the Kalb--Ramond field. The comparison is performed in transverse-traceless gauge, up to field redefinitions.
}

\begin{document}
\maketitle
\flushbottom

\section{The double copy as an alternative method to construct $\alpha'$-corrections}

\subsection{Motivation}

Higher-derivative corrections to string effective actions offer a way to explore whether the double copy program \cite{Kawai:1985xq,Bern:2008qj,Bern:2010ue,Monteiro:2014cda} can be used as a constructive tool for gravitational theories, potentially improving our understanding of quantum corrections to gravity.\footnote{For a pedagogical introduction to the off-shell double-copy program, see \cite{Lescano:2026xlb}.} This question has become particularly relevant at the level of actions, where recent results have shown that the quadratic and cubic contributions to perturbative Double Field Theory (DFT) \cite{Hull:2009mi} can be obtained from Yang--Mills theory through color-to-kinematics replacement rules \cite{Diaz-Jaramillo:2021wtl}, while extensions based on gauge-theory interactions containing $(DF)^2$ structures lead to higher-derivative, T-duality-invariant theories \cite{Lescano:2023pai,Lescano:2024gma,Lescano:2024lwn,Yilmaz:2024jvf}.

In particular, the construction of \cite{Lescano:2024lwn} established a connection between a higher-derivative gauge Lagrangian and the first $\alpha'$ corrections to the bosonic-string effective action at quadratic order in the fields, including the Kalb--Ramond and dilaton sectors. Extending this approach to order $\alpha'^2$ requires going beyond this quadratic approximation, since terms of the form $\mathrm{Riem}^3$ already contribute at cubic order in the gravitational perturbation. The corresponding six-derivative corrections are independently known \cite{Metsaev:1986yb,Metsaev:1987zx,Garousi:2019wgz,Ameri:2025bei}, making this order a controlled setting in which to test whether the double-copy construction can systematically determine both the required higher-derivative gauge-theory interactions and the associated kinematic replacement rules. At the level of on-shell amplitudes, higher-derivative operators and the double copy have been studied extensively, see e.g.~\cite{Broedel:2012rc,Johansson:2017srf,Johansson:2018ues,Huang:2016tag,Azevedo:2018dgo,Carrasco:2019yyn}; here we address the corresponding question directly at the level of the action and the replacement rules.

A natural next step is to ask whether the same strategy can be extended to the $\alpha'^3$ corrections. At this order, a particularly interesting curvature-quartic sector proportional to $\zeta(3)$ appears in string effective actions and presents obstructions to constructions based solely on the symmetry principles underlying duality-covariant formulations \cite{Hronek:2020xxi}. The present analysis can thus be viewed as a step toward determining whether an action-level double copy could provide an alternative route to such higher-order corrections.

\subsection{Strategy}

Our starting point is the off-shell double-copy construction of \cite{Diaz-Jaramillo:2021wtl}, in which the Yang--Mills action, up to cubic order in the fields, is mapped to perturbative DFT, whose pure gravitational sector reproduces Einstein--Hilbert gravity. The construction replaces color structures by kinematic ones and, at cubic order, the comparison with the gravitational action is performed in Siegel gauge. In momentum space, the Cartan--Killing metric and the structure constants are mapped according to
\begin{align}
\kappa_{ab} & \longrightarrow \frac{1}{2}\left(\eta^{\bar{\mu}\bar{\nu}} - \frac{\bar{k}^{\bar{\mu}}\bar{k}^{\bar{\nu}}}{\bar{k}^2}\right)\, ,
\label{idHohm1}
\\
f_{abc} & \longrightarrow \frac{i}{4}\left(\eta^{\bar{\mu}\bar{\nu}}\bar{k}_{12}^{\bar{\rho}} + \eta^{\bar{\nu}\bar{\rho}}\bar{k}_{23}^{\bar{\mu}} + \eta^{\bar{\rho}\bar{\mu}}\bar{k}_{31}^{\bar{\nu}}\right)\, ,
\label{idHohm2}
\end{align}
where $\bar{k}_{ij}=\bar{k}_{i}-\bar{k}_{j}$ and $\bar{\mu},\bar{\nu},\ldots$ denote a second set of spacetime indices. The gauge field is accordingly mapped to a doubled gravitational field $e_{\mu\bar \mu}$. After imposing the appropriate identifications (see section~\ref{Sec:review}), the field content comprises the metric, Kalb--Ramond field and dilaton degrees of freedom.

In this work, we focus on the bosonic-string effective action through order $\alpha'^2$. Throughout the explicit Lagrangians below, we set $\alpha'=1$ while retaining powers of $\alpha'$ as labels for the corresponding orders in the derivative expansion. Thus, the $\alpha'$ and $\alpha'^2$ corrections correspond to four- and six-derivative terms, respectively. We stress that a coefficient multiplying a term with $2n$ additional momenta carries a factor $\alpha'^n$; in particular, a deformation of the replacement rule \eqref{idHohm2} by terms cubic in the barred momenta is of order $\alpha'$, and contributes already at order $\alpha'$ when acting on the Yang--Mills vertex. Writing $S_{\mathrm{Bos}}=\int d^D x\,\sqrt{-g}\,e^{-2\Phi}L_{\mathrm{Bos}}$, the Lagrangian in the conventions used throughout this work is, in the scheme of \cite{Metsaev:1987zx,Ameri:2025bei},
\begin{align}
L_{\mathrm{Bos}} = {} & R + 4\partial_\mu\Phi\partial^\mu\Phi - \frac{1}{12}H_{\mu\nu\rho}H^{\mu\nu\rho} + \frac{1}{4}\Big(R_{\mu\nu\rho\lambda}R^{\mu\nu\rho\lambda} - \frac{1}{2}H_{\alpha}{}^{\delta\epsilon}H^{\alpha\beta\gamma}R_{\beta\gamma\delta\epsilon}\Big) \nn \\
& - \frac{1}{12}R^{\mu}{}_{\nu}{}^{\rho\lambda}R^{\nu}{}_{\sigma\lambda}{}^{\tau}R^{\sigma}{}_{\rho\tau\mu}
- \frac{1}{12}R^{\mu\nu}{}_{\rho\lambda}R^{\sigma}{}_{\tau}{}^{\rho\lambda}R^{\tau}{}_{\mu\sigma\nu} + \dots\, ,
\label{bosonicintro}
\end{align}
where $H_{\mu\nu\rho}=3\partial_{[\mu}b_{\nu\rho]}$ and the dots denote terms which are at least quartic in the fields, and therefore do not contribute to the cubic analysis of this work. In the scheme of \cite{Ameri:2025bei} the higher-derivative terms are multiplied by the same dilaton factor $e^{-2\Phi}$ as the two-derivative ones; since the analysis below is restricted to $\Phi=0$, this factor plays no role here. Since these higher-derivative corrections are known independently, their cubic expansion provides a useful benchmark for determining the gauge-theory interactions and kinematic replacement rules required by the action-level double copy. For this purpose, we work with perturbations around Minkowski space in the transverse-traceless (TT) gauge
\be
\partial^{\mu}h_{\mu\nu} = 0\, , \qquad h = 0\, , \qquad \partial^{\mu}b_{\mu\nu} = 0\, ,
\label{TT}
\ee
where $h=h_{\mu\nu}\eta^{\mu\nu}$. As we review in section~\ref{Sec:review}, the transversality of the Kalb--Ramond field is not an additional choice: it follows, together with the transversality of $h_{\mu\nu}$, from the Siegel-gauge conditions of the double copy, and it is required for compatibility with the two-derivative results of \cite{Hull:2009mi,Diaz-Jaramillo:2021wtl}. Throughout, the comparison between the double copy and the string effective action is performed in this gauge, up to integrations by parts and up to terms proportional to the lowest-order equations of motion, which can be removed by field redefinitions. The resulting statements are therefore statements about gauge-fixed cubic vertices, equivalent to on-shell three-point data (evaluated on complexified momenta, see appendix~\ref{App:support}); extending them to a fully gauge-invariant off-shell formulation of the deformed map is left for future work.

\subsection{Main results}

\begin{itemize}
\item We first determine the higher-derivative gauge-theory operators required to reproduce the gravitational corrections. We consider the deformation of the Yang--Mills Lagrangian
\begin{align}
L_{\mathrm{gauge}} = {} & - \frac{1}{4}\kappa_{ab}F_{\mu\nu}{}^{a}F^{\mu\nu b}
+ a_{1}\kappa_{ab}D^{\nu}F^{\rho\lambda a}D_{\rho}F_{\nu\lambda}{}^{b}
+ a_{2}f_{abc}F_{\mu\nu}{}^{a}F^{\nu\rho b}F_{\rho}{}^{\mu c}\, ,
\label{FinalLag}
\end{align}
where $D_{\mu}X^a = \partial_{\mu}X^a + g_{\mathrm{YM}}f^{a}{}_{bc}A_\mu{}^bX^c$ is the gauge-covariant derivative. We show that the $(DF)^2$ operator is related to the $F^3$ operator by an exact identity, up to total derivatives and terms proportional to the Maxwell equations. As a consequence, the cubic gravitational vertices depend only on the combination $\lambda = a_2 - g_{\mathrm{YM}}a_1$.

\item At order $\alpha'^2$, we show that the cubic structure emerging from the six-derivative gravitational Lagrangian cannot be generated by adding further higher-derivative terms to \eqref{FinalLag}, such as the $DDFDDF$ operators
\begin{align}
L = \beta_{1}\kappa_{ab}D^{\nu}D^{\rho}F^{\sigma\lambda a}D_{\sigma}D_{\rho}F_{\nu\lambda}{}^{b}
+ \beta_{2}\kappa_{ab}D^{\nu}D^{\rho}F^{\sigma\lambda a}D_{\nu}D_{\rho}F_{\sigma\lambda}{}^{b} + \dots
\label{Newbi}
\end{align}
More precisely, we establish two statements: no parity-even gauge-theory operator with more than three derivatives contributes to the cubic vertices in the gauge \eqref{TT}, and, with the standard replacement rule \eqref{idHohm2}, no gauge-theory Lagrangian whatsoever can produce the required index structure, in which every derivative index is contracted with a field index. Reproducing this six-derivative structure therefore requires extending the kinematic replacement of the structure constants by higher-derivative momentum terms associated with the $F^3$ vertex.

\item Our third result is the explicit construction of the required deformation of the kinematic replacement rule of \cite{Diaz-Jaramillo:2021wtl},
\be
f_{abc} \rightarrow \frac{i}{4} \bar \pi^{(3)}_{\bar \mu \bar \nu \bar \rho} + i c_{1}\Delta \bar \pi^{(3)}_{\bar \mu \bar \nu \bar \rho} \, ,
\label{fmap_intro}
\ee
where $\Delta\bar\pi^{(3)}$ is the three-momentum tensor of the $F^3$ vertex. Since $c_1$ is of order $\alpha'$, the deformation contributes both at order $\alpha'$ and at order $\alpha'^2$. The two resulting matching conditions for $(\lambda, c_1)$ admit a unique solution, which is a double root,
\be
\lambda = -\frac{1}{24}\, , \qquad c_1 = \frac{1}{8}\, ,
\ee
and for which the cubic action takes the left--right symmetric form $(\pi^{(3)}+\tfrac12\Delta\pi^{(3)})\otimes(\bar\pi^{(3)}+\tfrac12\Delta\bar\pi^{(3)})$. The gauge-theory deformation and the deformation of the map contribute equally: at order $\alpha'^2$, two distinct ingredients are required, and they are related by the exchange of the two copies. An asymmetric distribution of the same total four-derivative coupling, with no Riem$^3$ vertex, reproduces instead the heterotic pattern.

\item Finally, we extend the comparison to the Kalb--Ramond field. With $(\lambda,c_1)$ fixed in the pure gravitational sector, the double copy reproduces, without further input, the cubic Einstein--Hilbert and $H^2$ couplings, the cubic $\mathrm{Riem}\,H^2$ coupling of \eqref{bosonicintro} including its sign, the absence of cubic six-derivative couplings involving the Kalb--Ramond field, and the absence of cubic couplings odd in $b_{\mu\nu}$.
\end{itemize}

\section{Off-shell double copy of Yang--Mills theory in TT gauge}
\label{Sec:review}

In this section, we briefly review the construction of \cite{Diaz-Jaramillo:2021wtl} and specialize it to the gauge \eqref{TT}. We start from the $D$-dimensional Yang--Mills action
\be
\label{YM_action}
S_{\rm{YM}} = - \frac{1}{4}\int d^{D}x\,\kappa_{ab}F_{\mu \nu}{}^{a}F^{\mu\nu b}\, ,
\ee
where the Yang--Mills field strength is
\be
F_{\mu\nu}{}^{a} = 2\partial_{[\mu}A_{\nu]}{}^{a} + g_{\rm{YM}}f^{a}{}_{bc} A_{\mu}{}^{b}A_{\nu}{}^{c}\, ,
\ee
and spacetime indices are contracted with the Minkowski metric $\eta_{\mu\nu}=\mathrm{diag}(-,+,\ldots,+)$.

In momentum space, the quadratic Yang--Mills action takes the form
\be
\label{YM_momentum}
S_{\rm{YM}}^{(2)} = - \frac{1}{2}\int_{k}\,\kappa_{ab}\,k^2 \,\Pi^{\mu\nu}(k) A_{\mu}{}^{a}(-k) A_{\nu}{}^{b}(k)\; ,
\ee
where $\int_{k}\equiv\int d^Dk$ and
\be
\label{PiProjector}
\Pi^{\mu\nu}(k) = \eta^{\mu\nu}-\frac{k^{\mu}k^{\nu}}{k^2}\, .
\ee
The double-copy prescription replaces the color index by a second spacetime index, together with an independent momentum
\be
\label{DcFields}
A_{\mu}{}^{a}(k) \longrightarrow e_{\mu\bar{\mu}}(k,\bar{k})\, ,
\ee
while the Cartan--Killing metric is mapped according to
\be
\label{DCCartan}
\kappa_{ab} \longrightarrow \frac{1}{2}\bar{\Pi}^{\bar{\mu}\bar{\nu}}(\bar{k})\, ,
\ee
where $\bar{\Pi}^{\bar{\mu}\bar{\nu}}$ is defined analogously to \eqref{PiProjector} in terms of barred momenta and indices, reproducing the replacement rule \eqref{idHohm1}. The two momentum sectors are subject to the constraint
\be
\label{constraint}
\bar{k}^2 = k^{2}\, ,
\ee
which is the momentum-space counterpart of the level-matching condition of the doubled formulation. Applying these replacements to \eqref{YM_momentum} gives
\be
\label{quadraticDoubleCopy}
S_{\rm DC}^{(2)} = - \frac{1}{4}\int_{k,\bar{k}} k^2\,\Pi^{\mu\nu}(k)  \bar{\Pi}^{\bar{\mu}\bar{\nu}}(\bar{k}) e_{\mu\bar{\mu}}(-k,-\bar{k})e_{\nu\bar{\nu}}(k,\bar{k})\, .
\ee

Expanding the projectors and Fourier transforming to position space produces a nonlocal contribution. This can be localized by introducing an auxiliary scalar field,
\be
\phi = \frac{1}{k^2}k^{\mu}\bar{k}^{\bar{\nu}}e_{\mu\bar{\nu}}\, ,
\ee
whose elimination reproduces \eqref{quadraticDoubleCopy}. We then solve the constraint \eqref{constraint} by identifying the two coordinates, $x=\bar{x}$, and identify the doubled field and the scalar with the massless fields of the string,
\begin{align}
e_{\mu\bar{\nu}} & = \left(h_{\mu\nu}+b_{\mu\nu}\right)\delta^{\nu}{}_{\bar{\nu}}\, ,
\qquad
\phi = h - 4\Phi\, ,
\label{identification}
\end{align}
where $h_{\mu\nu}$ is the metric perturbation, $b_{\mu\nu}$ the Kalb--Ramond field and $\Phi$ the dilaton. With these identifications, the quadratic double-copy action agrees with the quadratic expansion of the two-derivative part of \eqref{bosonicintro} up to total derivatives.

At cubic order, the comparison is performed in Siegel gauge, in which the auxiliary vector fields of the formulation \cite{Hull:2009mi,Diaz-Jaramillo:2021wtl} are set to zero. Setting in addition the auxiliary scalars to zero, this implies
\be
\partial^{\mu}e_{\mu\bar\nu} = 0\, , \qquad \partial^{\bar\nu}e_{\mu\bar\nu} = 0\, , \qquad \phi = 0\, .
\label{Siegel}
\ee
Taking the sum and the difference of the first two conditions and using \eqref{identification}, one finds
\be
\partial^{\mu}h_{\mu\nu} = 0\, , \qquad \partial^{\mu}b_{\mu\nu} = 0\, , \qquad \Phi = \frac{1}{4}h\, .
\label{SiegelComponents}
\ee
Imposing only the transversality of $h_{\mu\nu}$ would instead mix $\partial^\mu h_{\mu\nu}$ with $\partial^\mu b_{\mu\nu}$, and the agreement with the two-derivative cubic action would be lost. Together with $h=0$, the conditions \eqref{SiegelComponents} give the gauge \eqref{TT} and $\Phi=0$. In other words, the analysis of this work is restricted to configurations with vanishing dilaton, and the dilaton couplings are not tested. Since we compare cubic vertices, this restriction simply means that the cubic couplings with an external dilaton are not computed. The $hhh$ and $hbb$ couplings analyzed below are not affected by it: at quadratic order the dilaton decouples from $h_{\mu\nu}$ in TT gauge, since it couples through $\Phi R^{(1)}(h)$ and $R^{(1)}(h)=\partial^\mu\partial^\nu h_{\mu\nu}-\Box h$ vanishes, and dilaton exchange would first affect the comparison at quartic order. In this gauge, the quadratic Lagrangian reduces to
\be
L_{\mathrm{DC}}^{(2)} = \frac{1}{4}h^{\mu\nu}\Box h_{\mu\nu} + \frac{1}{4}b^{\mu\nu}\Box b_{\mu\nu}\, ,
\ee
which agrees with the quadratic Einstein--Hilbert and $-\frac{1}{12}H^2$ Lagrangians in the same gauge. The previous Lagrangian shows that all the cubic terms proportional to $\Box h_{\mu \nu}$ or $\Box b_{\mu\nu}$ can be neglected since they can be eliminated by field redefinitions, which induce quartic terms.

We now turn to the cubic sector. In momentum space, the cubic Yang--Mills action can be written as
\be
\label{YM_cubic}
S^{(3)}_{\mathrm{YM}} = -\frac{i g_{\mathrm{YM}}}{6(2\pi)^{D/2}}\int_{k_i} \delta(k_i)f_{abc}\,\pi^{(3)\mu\nu\rho}A_{1\mu}^{a} A_{2\nu}^{b} A_{3\rho}^{c}\, ,
\ee
where
\be
\label{Pi_3index}
\pi^{(3)\mu\nu\rho}(k_1,k_2,k_3) = \eta^{\mu\nu} k_{12}^{\rho} + \eta^{\nu\rho} k_{23}^{\mu} + \eta^{\rho\mu} k_{31}^{\nu}\, ,
\ee
with $k_{ij}=k_i-k_j$ and $\delta(k_i)\equiv\delta(k_1+k_2+k_3)$. At this order, the double-copy prescription requires replacing the structure constants according to
\be
\label{DC_structure}
f_{abc} \longrightarrow \frac{i}{4} \bar{\pi}^{(3)\bar{\mu}\bar{\nu}\bar{\rho}}\, ,
\ee
which corresponds to the cubic replacement rule \eqref{idHohm2}. Applying the double-copy replacement rules to \eqref{YM_cubic} and setting $g_{\mathrm{YM}}=\frac{1}{2}$ gives
\be
\label{cubic}
S_{\mathrm{DC}}^{(3)} = \frac{1}{48(2\pi)^{D/2}}\int_{K_{i}}\delta(K_i)\,\bar{\pi}^{(3)\bar{\mu}\bar{\nu}\bar{\rho}}\pi^{(3)\mu\nu\rho}e_{1\mu\bar{\mu}}e_{2\nu\bar{\nu}}e_{3\rho\bar{\rho}}\, ,
\ee
where $K_{i}=(k_{i},\bar{k}_{i})$ and $e_{i\mu\bar{\mu}}=e_{\mu\bar{\mu}}(K_i)$. Fourier transforming to position space, integrating by parts, and imposing \eqref{TT}, we obtain
\be
\label{DC_cubic}
L^{(3)}_{\rm DC} = \frac{1}{4}e_{\mu\delta}\,\partial^{\mu}e_{\rho\sigma}\,\partial^{\delta}e^{\rho\sigma} - \frac{1}{4}e_{\mu\delta}\,\partial^{\mu}e_{\nu\rho}\,\partial^{\rho}e^{\nu\delta} - \frac{1}{4}e_{\mu\delta}\,\partial^{\rho}e^{\mu\nu}\,\partial^{\delta}e_{\rho\nu}\, ,
\ee
where, after the identification $x=\bar x$, the second index of $e_{\mu\nu}$ is the barred one. Terms odd in $b_{\mu\nu}$ cancel, and in the pure gravitational sector \eqref{DC_cubic} reduces to
\be
\label{DC_cubic_h}
L^{(3)}_{\rm DC}\big|_{h} = \frac{1}{4}\partial_{\mu}h_{\rho\lambda} \partial_{\nu}h^{\rho\lambda}h^{\mu\nu} - \frac{1}{2}\partial_{\mu}h^{\rho\nu}\partial_{\nu}h_{\rho\lambda}h^{\mu\lambda}\, .
\ee
Denoting by $L^{(3)}_{\mathrm{NS}}$ the cubic part of the two-derivative NS--NS Lagrangian, $\sqrt{-g}\,e^{-2\Phi}(R+4\partial_\mu\Phi\partial^\mu\Phi-\frac{1}{12}H^2)$, in the same gauge, one finds
\be
\label{EH_cubic}
L^{(3)}_{\mathrm{NS}} \simeq L^{(3)}_{\rm DC} - \frac{1}{4}\Box e^{\nu}{}_{\mu}\,e_{\nu}{}^{\rho}\,e_{\rho}{}^{\mu}\, ,
\ee
where $\simeq$ denotes equality up to total derivatives. The last term is proportional to the linearized equations of motion and can therefore be removed by the field redefinition $e_{\mu\nu}\to e_{\mu\nu}-\frac{1}{2}e_{\mu\rho}e^{\rho}{}_{\nu}$, establishing the equivalence of the two cubic Lagrangians.

It is useful to note that, in the gauge \eqref{TT} and up to field redefinitions, a cubic Lagrangian is completely characterized by its value on three plane waves with null momenta satisfying $k_i\cdot k_j=0$ and polarizations transverse on both indices. On this support, each structure reduces to a polynomial in two Lorentz-invariant building blocks for each copy, $\cA_{\mathrm{YM}}$ and $\cA_{F^3}$, defined in appendix~\ref{App:support}. This provides a simple and systematic way to perform the integrations by parts and field redefinitions below, and we will use the notation $\cong$ for equality on this support. In particular, on the support
\be
\pi^{(3)\mu\nu\rho}\cong 2\,T_{\mathrm{YM}}^{\mu\nu\rho}\, , \qquad
\Delta\pi^{(3)\mu\nu\rho} \cong 2\,T_{F^3}^{\mu\nu\rho}\, ,
\label{supportPi}
\ee
where $T_{\mathrm{YM}}$ and $T_{F^3}$ are defined in \eqref{Ttensors} and $\Delta\pi^{(3)}$ will be introduced in section~\ref{Sec:alpha2}.

\section{Double copy at order $\alpha'$}
\label{Sec:alpha1}

The first higher-derivative correction to the bosonic-string effective Lagrangian in the pure gravitational sector is given by
\be
L^{\alpha'}_{\mathrm{Bos}} = \frac{1}{4}R_{\mu\nu\rho\lambda}R^{\mu\nu\rho\lambda}\, .
\ee
At quadratic order in the gravitational perturbation, the curvature-squared term reduces in TT gauge, up to total derivatives, to
\be
\frac{1}{4}\left.R_{\mu\nu\rho\lambda}R^{\mu\nu\rho\lambda}\right|_{\mathrm{TT}}^{(2)} = \frac{1}{4}\Box h_{\mu\nu}\Box h^{\mu\nu}\, ,
\ee
where we used that the quadratic part of the Gauss--Bonnet combination is a total derivative and that the linearized Ricci tensor reduces to $-\frac{1}{2}\Box h_{\mu\nu}$. This contribution is proportional to the lowest-order equations of motion and can be removed by a field redefinition. Consequently, no additional four-derivative gauge-theory contribution is required at quadratic order.

At cubic order in the gravitational perturbation, the curvature-squared term reduces in TT gauge, after integrations by parts and neglecting terms that can be removed by field redefinitions, to
\be
\frac{1}{4}\left.R_{\mu\nu\rho\lambda}R^{\mu\nu\rho\lambda}\right|_{\mathrm{TT}}^{(3)} \simeq \frac{1}{4}\,\mathcal{S}_1\, ,
\qquad
\mathcal{S}_1 \equiv \partial_\mu\partial_\nu h^\rho{}_\lambda\, \partial_\sigma\partial_\rho h^{\lambda\mu}\,h^{\nu\sigma}\, .
\label{Riem23}
\ee
A second structure, $\partial_\mu\partial_\nu h^{\rho}{}_{\lambda}\partial_\sigma\partial^\mu h^{\lambda}{}_{\rho}h^{\nu\sigma}$, can appear in intermediate steps. Since the index $\mu$ is contracted between derivatives acting on two different fields, integrating by parts turns it into terms with a $\Box$ acting on a single field, and it is therefore removable by field redefinitions. Unlike the quadratic contribution, $\mathcal{S}_1$ cannot be eliminated by field redefinitions and must therefore be reproduced by higher-derivative interactions on the gauge-theory side. We consider the ansatz
\be
L^{\alpha'}_{\rm gauge} = a_1 \kappa_{a b}D^{\nu}{F^{\sigma \epsilon a}} D_{\sigma}{F_{\nu \epsilon}{}^{b}} + a_2 f_{a b c} F_{\mu \nu}{}^{a} F^{\nu \rho b} F_{\rho}{}^{\mu c} \, .
\label{Lhigh}
\ee

\paragraph{An identity for the $(DF)^2$ operator.}
Integrating by parts once and using $[D_\mu,D_\nu]X^a=g_{\mathrm{YM}}f^a{}_{bc}F_{\mu\nu}{}^bX^c$, one finds the exact identity
\be
\kappa_{ab}D^{\nu}F^{\sigma\epsilon a}D_{\sigma}F_{\nu\epsilon}{}^{b} = \kappa_{ab}D_{\sigma}F^{\sigma\epsilon a}D^{\nu}F_{\nu\epsilon}{}^{b} - g_{\mathrm{YM}}\,f_{abc}F_{\mu\nu}{}^{a}F^{\nu\rho b}F_{\rho}{}^{\mu c} + \partial_\mu(\dots)\, .
\label{DFidentity}
\ee
The first term on the right-hand side is quadratic in the Yang--Mills equations of motion and can be removed by a field redefinition of the gauge field. Therefore, up to field redefinitions, the ansatz \eqref{Lhigh} is equivalent to a single $F^3$ operator with coefficient
\be
\lambda \equiv a_2 - g_{\mathrm{YM}}\,a_1\, .
\label{lambda}
\ee
This is reflected in the cubic gravitational vertices below, which depend on $a_1$ and $a_2$ only through $\lambda$. In particular, the index contraction of the $(DF)^2$ operator in \eqref{Lhigh} differs from the one of \cite{Lescano:2024lwn}, $\kappa_{ab}D_\mu F^{\mu\nu a}D_\rho F^{\rho}{}_\nu{}^b$, only by an $F^3$ term.\footnote{The operator of \cite{Lescano:2024lwn} is the first term on the right-hand side of \eqref{DFidentity}. It is quadratic in the Yang--Mills equations of motion and therefore does not contribute to the cubic vertices considered here, up to field redefinitions.}

\paragraph{Cubic vertex.}
Since the quadratic contribution does not affect the gravitational theory up to field redefinitions, we can focus directly on the cubic order. On the support described in section~\ref{Sec:review}, and setting $g_{\mathrm{YM}}=\frac{1}{2}$, the complete cubic vertex of $L_{\mathrm{YM}}+L^{\alpha'}_{\mathrm{gauge}}$ takes the compact form
\be
S^{(3)}_{\mathrm{gauge}} \cong -\frac{i}{12(2\pi)^{D/2}}\int_{k_i}\delta(k_i)\,f_{abc}\left(\pi^{(3)\mu\nu\rho}-12\lambda\,\Delta\pi^{(3)\mu\nu\rho}\right)A_{1\mu}^{a}A_{2\nu}^{b}A_{3\rho}^{c}\, ,
\label{gaugeVertex}
\ee
where
\be
\Delta\pi^{(3)\mu\nu\rho} = k_1^{\rho}k_2^{\mu}k_3^{\nu} - k_1^{\nu}k_3^{\mu}k_2^{\rho}
\label{DeltaPiUnbarred}
\ee
is the three-momentum tensor of the $F^3$ vertex. All other terms in the momentum-space vertex of \eqref{Lhigh} contain at least one factor $k_i\cdot k_j$ or $k_i\cdot A_i$ and are removable. Applying the replacement $f_{a b c}\rightarrow\frac{i}{4}\bar{\pi}^{(3)\bar{\mu}\bar{\nu}\bar{\rho}}$, the resulting gravitational contribution reduces, after integrations by parts and neglecting terms removable by field redefinitions, to
\be
L^{\alpha'(3)}_{\mathrm{DC}} \simeq -3\lambda\,\mathcal{S}_1\, .
\label{alpha1standard}
\ee
Matching this result with the cubic curvature-squared correction \eqref{Riem23} fixes, with the standard replacement rule,
\be
\lambda = - \frac{1}{12}\, .
\label{lambdaStandard}
\ee
As we show in the next section, this value is modified once the replacement rule is deformed, since the deformation also contributes at order $\alpha'$.

\section{Double copy at order $\alpha'^2$}
\label{Sec:alpha2}

We now extend the construction to the six-derivative correction of the bosonic-string effective action. In the pure gravitational sector, the order-$\alpha'^2$ contribution is
\begin{align}
L_{\mathrm{Bos}}^{\alpha'^2} = {} & - \frac{1}{12}R^{\mu}{}_{\nu}{}^{\rho\lambda}R^{\nu}{}_{\sigma\lambda}{}^{\tau}R^{\sigma}{}_{\rho\tau\mu}
- \frac{1}{12}R^{\mu\nu}{}_{\rho\lambda}R^{\sigma}{}_{\tau}{}^{\rho\lambda}R^{\tau}{}_{\mu\sigma\nu}\, .
\end{align}
At cubic order in the gravitational perturbation, each curvature can be replaced by its linearized expression, and this contribution reduces in TT gauge, after integrations by parts and neglecting terms removable by field redefinitions, to
\be
L_{\mathrm{Bos}}^{\alpha'^2(3)}\big|_{\rm TT} \simeq - \frac{1}{48}\,\partial^\mu\partial^\nu h^{\rho\lambda}\,\partial_\rho\partial_\lambda h^{\sigma\tau}\,\partial_\sigma\partial_\tau h_{\mu\nu}
\equiv - \frac{1}{48}\, \mathcal O_{1}\, .
\label{Riem3cubicterms}
\ee
Separately, the two Riem$^3$ invariants reduce to $-\frac{1}{4}\mathcal O_1$ and $\frac{1}{2}\mathcal O_1$, respectively. This reflects the fact that a single independent six-derivative three-graviton structure survives on this support. The structure $\mathcal{O}_1$ is distinguished by the fact that every derivative index is contracted with a field index.

\subsection{Absence of a gauge-theory origin with the standard replacement rule}

We now show that $\mathcal{O}_1$ cannot be generated by adding further higher-derivative operators to the gauge-theory Lagrangian while keeping the replacement rule \eqref{DC_structure}. The argument relies on two simple observations about cubic vertices on the support of section~\ref{Sec:review}.

\paragraph{(i) Operators with more than three derivatives do not contribute.} A cubic gauge-theory vertex is trilinear in the polarizations $\varepsilon_i$ of the three legs. On the support, $\varepsilon_i\cdot k_i=0$ and $k_i\cdot k_j=0$, and momentum conservation allows one to express each $k_i$ contracted with $\varepsilon_j$ in terms of a single momentum. Every momentum must therefore be contracted with a polarization, and with three polarizations at most three momenta survive. Hence any cubic vertex with five or more momenta vanishes up to integrations by parts and field redefinitions. This applies in particular to the $DDFDDF$ operators \eqref{Newbi}, and to any other operator of the schematic form $D^nF D^nF$ with $n\geq 2$, which therefore do not introduce further cubic terms. Moreover, on the support only two independent parity-even structures survive, $\cA_{\mathrm{YM}}$ with one momentum and $\cA_{F^3}$ with three, so that the cubic sector of any parity-even gauge-theory Lagrangian built from $\kappa_{ab}$ and $f_{abc}$ is characterized, in generic dimension, by a single higher-derivative coupling, which is $\lambda$ in \eqref{gaugeVertex}.

\paragraph{(ii) The standard rule always produces contracted field indices.} With the replacement \eqref{DC_structure}, the barred side of every cubic gravitational vertex is $\bar\pi^{(3)\bar\mu\bar\nu\bar\rho}$, which contains the barred metric in each of its terms. After the identification $x=\bar x$, this metric contracts two field indices of the gravitational perturbations. On the contrary, in $\mathcal{O}_1$ every field index is contracted with a derivative, and on the support $\mathcal{O}_1 \cong -6\,\cA_{F^3}\bar\cA_{F^3}$, with no factor containing the barred metric. Therefore $\mathcal{O}_1$ cannot be produced from any gauge-theory Lagrangian with the standard replacement rule, irrespective of the basis of gauge-theory operators.

The standard replacement rule is therefore insufficient to reproduce the complete cubic six-derivative gravitational Lagrangian, and we instead extend the kinematic replacement of the structure constants.

\subsection{Deformation of the replacement rule}

We deform the cubic rule \eqref{DC_structure} according to
\be
f_{abc}\longrightarrow\frac{i}{4}\bar{\pi}^{(3) \bar{\mu}\bar{\nu}\bar{\rho}} + i c_1 \Delta\bar{\pi}^{(3)\bar{\mu}\bar{\nu}\bar{\rho}}\, .
\label{fmap}
\ee
The deformation must contain three barred momenta, so that the barred metric is absent, see observation (ii). Higher-order deformations would instead generate interactions beyond the derivative order considered here, or vanish on the support by observation (i). We consider
\be
\Delta\bar{\pi}^{(3)}_{\bar{\mu}\bar{\nu}\bar{\rho}} = \bar{k}_{1\bar{\rho}}\bar{k}_{2\bar{\mu}}\bar{k}_{3\bar{\nu}} - \bar{k}_{1\bar{\nu}}\bar{k}_{3\bar{\mu}}\bar{k}_{2\bar{\rho}}\, .
\label{DeltaPi}
\ee
This tensor is totally antisymmetric under simultaneous permutations of the barred indices $(\bar{\mu},\bar{\nu},\bar{\rho})$ and the external legs $(1,2,3)$, in the same way as $\bar{\pi}^{(3)}_{\bar{\mu}\bar{\nu}\bar{\rho}}$. It coincides with the barred version of the three-momentum tensor \eqref{DeltaPiUnbarred} of the vertex generated by
\be
f_{abc}F_{\mu\nu}{}^aF^{\nu\rho b}F_\rho{}^{\mu c}\, ,
\ee
which is already present in the order-$\alpha'$ gauge-theory Lagrangian \eqref{Lhigh}.\footnote{At tree level, this interaction was also studied in \cite{Garozzo:2018uzj}.} Under the deformed double-copy map, this produces terms in which each momentum index is contracted with a field index, precisely the pattern required to generate $\mathcal O_1$.

Since $\Delta\bar\pi^{(3)}$ carries two more momenta than $\bar\pi^{(3)}$, the coefficient $c_1$ is of order $\alpha'$. Applying \eqref{fmap} to the complete cubic vertex \eqref{gaugeVertex} therefore produces two contributions linear in $c_1$: the action of $\Delta\bar\pi^{(3)}$ on the Yang--Mills vertex, which contributes at order $\alpha'$, and its action on the $F^3$ vertex, which contributes at order $\alpha'^2$. The complete cubic double-copy action is
\be
S^{(3)}_{\mathrm{DC}} \cong \frac{1}{48(2\pi)^{D/2}}\int_{K_i}\delta(K_i)\left(\bar\pi^{(3)\bar\mu\bar\nu\bar\rho}+4c_1\Delta\bar\pi^{(3)\bar\mu\bar\nu\bar\rho}\right)\left(\pi^{(3)\mu\nu\rho}-12\lambda\,\Delta\pi^{(3)\mu\nu\rho}\right)e_{1\mu\bar\mu}e_{2\nu\bar\nu}e_{3\rho\bar\rho}\, ,
\label{DCdeformed}
\ee
whose pure gravitational part reduces, after integrations by parts and neglecting terms removable by field redefinitions, to
\be
L^{(3)}_{\mathrm{DC}} \simeq L^{(3)}_{\mathrm{DC}}\big|_{h} + \left(c_1-3\lambda\right)\mathcal{S}_1 + 4c_1\lambda\,\mathcal{O}_1\, .
\label{DeltaContributionGeneral}
\ee
Comparing with \eqref{Riem23} and \eqref{Riem3cubicterms}, the matching conditions at orders $\alpha'$ and $\alpha'^2$ read
\be
c_1 - 3\lambda = \frac{1}{4}\, ,\qquad 4c_1\lambda = -\frac{1}{48}\, .
\label{matching}
\ee
Eliminating $c_1$, one finds $(24\lambda+1)^2=0$. The system therefore has a unique solution, which is a double root,
\be
\lambda = -\frac{1}{24}\, ,\qquad c_1 = \frac{1}{8}\, .
\label{solution}
\ee
For instance, one may take $a_1=0$ and $a_2=-\frac{1}{24}$; any other choice with $a_2-\frac{1}{2}a_1=-\frac{1}{24}$ is equivalent up to field redefinitions. With these values, the gauge-theory deformation and the deformation of the map contribute equally to the order-$\alpha'$ vertex, $-3\lambda=c_1=\frac{1}{8}$, and the cubic action \eqref{DCdeformed} becomes
\be
S^{(3)}_{\mathrm{DC}} \cong \frac{1}{48(2\pi)^{D/2}}\int_{K_i}\delta(K_i)\left(\bar\pi^{(3)\bar\mu\bar\nu\bar\rho}+\tfrac{1}{2}\Delta\bar\pi^{(3)\bar\mu\bar\nu\bar\rho}\right)\left(\pi^{(3)\mu\nu\rho}+\tfrac{1}{2}\Delta\pi^{(3)\mu\nu\rho}\right)e_{1\mu\bar\mu}e_{2\nu\bar\nu}e_{3\rho\bar\rho}\, ,
\label{symmetric}
\ee
which is symmetric under the exchange of barred and unbarred quantities. In this sense, the cubic sector of the bosonic string through order $\alpha'^2$ is the double copy of $\mathrm{YM}+F^3$ with itself, where one copy is realized through the gauge-theory Lagrangian and the other through the replacement rule. Thus, unlike the order-$\alpha'$ construction with the standard map, reproducing the complete cubic $\alpha'^2$ correction requires both a deformation of the gauge-theory dynamics and a higher-derivative deformation of the kinematic replacement rule, and the two are related by the exchange of the two copies.

The double-root structure has a simple interpretation. Writing $u\equiv-12\lambda$ and $v\equiv4c_1$, the cubic gravitational action depends on the two copies through $u+v$ and $uv$. A target theory with Lagrangian $R+\beta\,\mathrm{Riem}^2+\dots$ and with a cubic six-derivative vertex $-\gamma\,\mathcal{O}_1$ leads to $u+v=4\beta$ and $uv=12\gamma$, whose two solutions are related by the exchange $u\leftrightarrow v$. For the bosonic string, $\beta=\frac14$ and $\gamma=\frac{1}{48}$, so that the two solutions coincide, $u=v=\frac{1}{2}$, and the construction is forced to be symmetric. For a theory with $\beta=\frac{1}{8}$ and $\gamma=0$, as in the pure gravitational sector of the heterotic string at this order, one finds instead $(u,v)=(\frac{1}{2},0)$ or $(0,\frac{1}{2})$: the $F^3$ deformation appears in only one of the two copies, either in the gauge-theory Lagrangian with the standard replacement rule, or in the replacement rule with the pure Yang--Mills Lagrangian. This mirrors the known pattern of three-point string amplitudes, $(\mathrm{YM}+F^3)\otimes(\mathrm{YM}+F^3)$ for the bosonic string and $\mathrm{YM}\otimes(\mathrm{YM}+F^3)$ for the heterotic string \cite{Kawai:1985xq,Huang:2016tag,Azevedo:2018dgo}, and it parallels the bosonic/heterotic dichotomy of the generalized Bergshoeff--de Roo deformations in duality-covariant formulations \cite{Bergshoeff:1989de,Marques:2015vua}. We return to the asymmetric case in section~\ref{Sec:B}.

\section{The Kalb--Ramond sector}
\label{Sec:B}

With $\lambda$ and $c_1$ fixed in the pure gravitational sector, the action \eqref{symmetric} contains no free parameters, and its Kalb--Ramond couplings are a prediction of the construction. In this section we compare them with the bosonic-string Lagrangian \eqref{bosonicintro}. We use the identification \eqref{identification} and the gauge \eqref{TT}, which in particular requires $\partial^\mu b_{\mu\nu}=0$, and in which $\Phi=0$.

Let us first clarify what is being tested. At three points, the closed-string vertex factorizes into the product of two open-string vertices, so that, once the three-graviton vertex is of the form $(\mathrm{YM}+F^3)\otimes(\mathrm{YM}+F^3)$, the three-point couplings of the Kalb--Ramond field are fixed by the KLT relations \cite{Kawai:1985xq}. The nontrivial content of the present section is that the action-level map, with the identification $e_{\mu\bar\nu}=h_{\mu\nu}+b_{\mu\nu}$, the Siegel-gauge conditions \eqref{SiegelComponents} and the normalization inherited from the two-derivative construction, lands on this factorized form without assuming it: the relative normalization of the two copies and the relative sign of the gauge-theory and map deformations are inputs of the construction, and both are tested by the Kalb--Ramond couplings. As shown below, the $hbb$ coupling at order $\alpha'$ is sensitive to the total four-derivative coupling $u+v$ and to the identification of $b_{\mu\nu}$ with the antisymmetric part of $e_{\mu\bar\nu}$, while the double-root structure itself is tested only by the three-graviton vertex.

At cubic order and in this gauge, the relevant terms of \eqref{bosonicintro} are
\begin{align}
L^{(3)}_{\mathrm{Bos}}\big|_{hbb} \simeq {} & \Big[\sqrt{-g}\,\Big(R-\frac{1}{12}H_{\mu\nu\rho}H^{\mu\nu\rho}\Big)\Big]^{(3)}_{hbb}
- \frac{1}{8}\,R^{(1)}_{\beta\gamma\delta\epsilon}(h)\,H_{\alpha}{}^{\delta\epsilon}H^{\alpha\beta\gamma}\, ,
\label{Bcubic}
\end{align}
where $R^{(1)}(h)$ is the linearized Riemann tensor and the first term reduces to $\frac{1}{4}h^{\mu\alpha}H_{\mu\nu\rho}H_{\alpha}{}^{\nu\rho}$, since $h=0$ and $\Phi=0$. At order $\alpha'^2$, all the terms of the bosonic-string Lagrangian involving the Kalb--Ramond field are at least quartic in the fields \cite{Garousi:2019wgz,Ameri:2025bei}, so that no cubic six-derivative coupling involving $b_{\mu\nu}$ is expected.\footnote{This statement refers to the scheme of \cite{Garousi:2019wgz,Ameri:2025bei}. Since the cubic vertices considered here are invariant under field redefinitions, the same conclusion holds in any scheme in which cubic six-derivative terms involving $b_{\mu\nu}$ are present, such terms being removable.} Similarly, the action is invariant under $b_{\mu\nu}\to-b_{\mu\nu}$, so that there are no couplings of the form $hhb$ or $bbb$.

On the double-copy side, these features follow from \eqref{symmetric}:
\begin{itemize}
\item \emph{Absence of couplings odd in $b_{\mu\nu}$.} The exchange of barred and unbarred quantities acts on the doubled field as $e_{\mu\bar\nu}\to e_{\nu\bar\mu}$, that is, $h_{\mu\nu}\to h_{\mu\nu}$ and $b_{\mu\nu}\to-b_{\mu\nu}$. Since \eqref{symmetric} is invariant under this exchange, all cubic terms odd in $b_{\mu\nu}$ vanish, to all orders in $\alpha'$ at cubic order. This is the cubic counterpart of the $\mathbb{Z}_2$ symmetry of the bosonic-string action.

\item \emph{Two-derivative couplings.} The two-derivative part of \eqref{symmetric} is \eqref{cubic}, which reproduces the $hbb$ couplings of the first term in \eqref{Bcubic}, up to the field redefinition discussed below \eqref{EH_cubic}.

\item \emph{Order $\alpha'$.} The order-$\alpha'$ part of \eqref{symmetric} is $\frac{1}{2}(\bar\pi^{(3)}\,\Delta\pi^{(3)}+\Delta\bar\pi^{(3)}\,\pi^{(3)})$. Since the exchange of the two copies maps $b_{\mu\nu}\to-b_{\mu\nu}$, the two terms give identical $hbb$ contributions, so that the $hbb$ coupling depends only on the total four-derivative coupling $u+v$ of section~\ref{Sec:alpha2}, which is fixed by the coefficient of Riem$^2$. Evaluated on $e=h+b$, the resulting $hbb$ vertex reproduces exactly the $\mathrm{Riem}\,H^2$ coupling in \eqref{Bcubic}, including its coefficient and sign. This is a nontrivial check, since no parameter is left after the pure gravitational matching: a $\mathrm{Riem}\,H^2$ coupling with any other coefficient, in particular with the opposite sign, would not be reproduced.

\item \emph{Order $\alpha'^2$.} The order-$\alpha'^2$ part of \eqref{symmetric} is $\frac14\Delta\bar\pi^{(3)}\,\Delta\pi^{(3)}$. By \eqref{supportPi}, on the support it is proportional to
\be
\big(k_{2}^{\mu}k_{2}^{\bar\mu}e_{1\mu\bar\mu}\big)\big(k_{3}^{\nu}k_{3}^{\bar\nu}e_{2\nu\bar\nu}\big)\big(k_{1}^{\rho}k_{1}^{\bar\rho}e_{3\rho\bar\rho}\big)\, ,
\ee
which vanishes as soon as one of the fields is antisymmetric. Hence the double copy predicts no cubic six-derivative couplings involving the Kalb--Ramond field, in agreement with the bosonic string. The order-$\alpha'^2$ deformation therefore affects only the three-graviton vertex, where it generates $\mathcal{O}_1$.
\end{itemize}
To summarize, with the values \eqref{solution} the cubic double-copy action reproduces
\be
S^{(3)}_{\mathrm{DC}} \simeq S^{(3)}_{\mathrm{Bos}}\big|_{\Phi=0}\, ,
\ee
in the gauge \eqref{TT}, for all couplings among $h_{\mu\nu}$ and $b_{\mu\nu}$, up to field redefinitions. We have verified these statements independently by evaluating the cubic expansion of \eqref{bosonicintro} and the double-copy action on the support described in appendix~\ref{App:support}, for arbitrary transverse polarizations.

The asymmetric solutions discussed at the end of section~\ref{Sec:alpha2} behave differently. They give the same $hbb$ coupling at order $\alpha'$ for the same value of $u+v$, but, since they break the exchange symmetry, they also generate at order $\alpha'$ a nonvanishing $hhb$ coupling. Indeed, the two terms of the order-$\alpha'$ vertex, $\bar\pi^{(3)}\,\Delta\pi^{(3)}$ and $\Delta\bar\pi^{(3)}\,\pi^{(3)}$, are exchanged by $b_{\mu\nu}\to-b_{\mu\nu}$ and therefore give opposite $hhb$ contributions, so that the $hhb$ coupling is proportional to $u-v$ (see appendix~\ref{App:support}). This is the cubic coupling expected from the Lorentz Chern--Simons modification of the three-form field strength in the heterotic string, schematically $H\to H+\alpha'\,\omega_{\mathrm{L}}$ with a convention-dependent coefficient, which contributes a term $H\cdot\omega_{\mathrm{L}}$ quadratic in $h_{\mu\nu}$ \cite{Bergshoeff:1989de}. A detailed comparison with the heterotic effective action, including its gauge sector, is beyond the scope of this work.

\section{Discussion and conclusions}

In this work, we have developed an action-level double-copy construction for the bosonic-string effective action through order $\alpha'^2$ and cubic order in the metric and Kalb--Ramond perturbations. Starting from the double-copy prescription of \cite{Diaz-Jaramillo:2021wtl}, we introduced the higher-derivative, gauge-invariant deformation of Yang--Mills theory given in \eqref{FinalLag}, and showed that, up to field redefinitions, it is controlled by a single $F^3$ coupling. We showed that no gauge-theory Lagrangian can reproduce the cubic Riem$^3$ vertex with the standard replacement rule, and that the required deformation of the replacement of the structure constants is the three-momentum tensor of the $F^3$ vertex. Since this deformation is of order $\alpha'$, it contributes both to the curvature-squared and to the curvature-cubic vertices, and the two resulting matching conditions admit a unique solution, a double root, for which the cubic action is the left--right symmetric product $(\mathrm{YM}+F^3)\otimes(\mathrm{YM}+F^3)$. In this sense, the gravitational higher-derivative expansion is encoded not only in the gauge-theory dynamics, but also in the color-to-kinematics map itself, and the two are related by the exchange of the two copies.

The construction has no free parameters once the pure gravitational sector is matched, and its extension to the Kalb--Ramond field provides a nontrivial test. With the identification $e_{\mu\bar\nu}=h_{\mu\nu}+b_{\mu\nu}$ and the transversality of both fields implied by Siegel gauge, the double copy reproduces the cubic $\mathrm{Riem}\,H^2$ coupling of the bosonic string with the correct coefficient and sign, and it predicts the absence of cubic six-derivative couplings involving the Kalb--Ramond field and of couplings odd in $b_{\mu\nu}$, in agreement with the string effective action. The asymmetric distribution of the $F^3$ deformation, in which only one of the two copies is deformed, reproduces instead the pure gravitational pattern of the heterotic string and generates the $hhb$ coupling associated with the Lorentz Chern--Simons term.

The symmetric form \eqref{symmetric} also gives a more precise meaning to the relation between the replacement rules and the gauge-theory kinematics. On the support, the deformed rule \eqref{fmap} with $c_1=\frac18$ replaces the structure constants by the barred copy of the color-stripped cubic vertex of the deformed gauge theory itself, $\pi^{(3)}+\frac12\Delta\pi^{(3)}$. This suggests interpreting $\bar\pi^{(3)}+\frac12\Delta\bar\pi^{(3)}$ as an $F^3$ deformation of the kinematic structure constants of Yang--Mills theory, and the order-$\alpha'$ and order-$\alpha'^2$ cubic vertices as the cross and diagonal terms of the product of two such deformed kinematic structures, even though our prescription is formulated as a replacement of color tensors and fields rather than as a tensor-product construction.

A useful feature of the present construction is that the bosonic-string effective action \eqref{bosonicintro} provides an independently known benchmark against which the double-copy procedure can be tested. This allows us to identify how the higher-derivative information is distributed between the gauge-theory interactions and the kinematic replacement rules. Beyond this regime, the double-copy construction could instead be constrained directly by the symmetries of the string effective theory. In particular, the tree-level effective action restricted to backgrounds with $d$ commuting isometries admits an $O(d,d)$ symmetry to all orders in $\alpha'$ \cite{Sen:1991zi,Hohm:2014sxa}. Requiring compatibility with T-duality, together with diffeomorphism invariance and the appropriate additional symmetries of the theory, may therefore provide independent constraints on higher-order gauge-theory interactions and deformations of the kinematic map when the gravitational target is not known in advance.

It is also useful to compare the present construction with \cite{Lescano:2024lwn}, where the quadratic higher-derivative corrections to the bosonic-string effective action, including the Kalb--Ramond and dilaton sectors, were obtained from an off-shell double-copy map. That construction was based on the higher-derivative gauge theory introduced in \cite{Johansson:2017srf,Johansson:2018ues} and related to the bosonic sectorized string in \cite{Azevedo:2019zbn}. In the sectors considered here, and after imposing the gauge \eqref{TT}, the contributions associated with $\kappa_{ab}D_\mu F^{\mu\nu a}D_\rho F^\rho{}_{\nu}{}^b$ are removable by field redefinitions, and, by the identity \eqref{DFidentity}, the $(DF)^2$ contraction of \eqref{Lhigh} contributes to the cubic vertices only through its $F^3$ component. It would be interesting to understand the role of the $(DF)^2$ operators of \cite{Johansson:2017srf,Johansson:2018ues} at quartic order, where the identity \eqref{DFidentity} no longer removes their contributions.

\subsection{Outlook}

The results of this work suggest several continuations. The most promising one is the extension of the construction to quartic order in the fields, with the aim of constructing the curvature-quartic sector proportional to $\zeta(3)$ that appears at order $\alpha'^3$. Since the leading perturbative contribution of a $\mathrm{Riem}^4$ invariant is quartic in $h_{\mu\nu}$, and since all cubic vertices are exhausted by the structures $\cA_{\mathrm{YM}}$ and $\cA_{F^3}$ of each copy, no $\zeta(3)$ information is accessible at cubic order.\footnote{Off-shell and gauge-invariant double-copy constructions to quartic order have been developed using homotopy-algebraic methods \cite{Bonezzi:2022bse}. Related BRST and homotopy-algebraic formulations of the Lagrangian double copy can be found in \cite{Borsten:2021hua,Borsten:2023ned}.} The present results suggest a concrete strategy. The cubic replacement rule \eqref{symmetric} has to be supplemented by quartic replacement rules for products of structure constants, compatible with the deformed cubic rule and with the Jacobi identity of the deformed kinematic structure constants. At the same time, the gauge-theory Lagrangian acquires new independent operators at quartic order, of the schematic form $F^4$ and $(DF)^2F$, which are no longer removable, and the left--right symmetry found here suggests that each of them should be accompanied by a corresponding deformation of the quartic replacement rules. Finally, the closed-string amplitudes are related to the open-string ones through the single-valued projection \cite{Stieberger:2013wea,Schlotterer:2012ny}, under which the $\zeta(2)$ contributions of the open string drop out and the $\zeta(3)$ contributions are doubled. An action-level realization of this projection, in which the $\zeta(2)$ terms cancel between the gauge-theory Lagrangian and the replacement rules while the $\zeta(3)$ terms add up, would provide a natural mechanism for generating the $\zeta(3)$ sector. These contributions are of particular interest because they encounter obstructions in duality-covariant constructions \cite{Hronek:2020xxi,Hsia:2024kpi}, and they therefore provide a natural testing ground for the constructive potential of the action-level double copy.

A second natural continuation concerns the algebraic structure underlying the replacement rules. The color-to-kinematics replacements employed in this work are defined at the level of the quadratic and cubic actions, and they are not assumed to be an algebra homomorphism. The natural framework in which to make the interpretation of the deformed rule as a deformed kinematic algebra precise is that of homotopy Lie ($L_\infty$) algebras \cite{Bonezzi:2022bse,Borsten:2021hua}, in which a gauge theory is described by an action $S=\frac12\langle\Psi,B_1(\Psi)\rangle+\frac1{3!}\langle\Psi,B_2(\Psi,\Psi)\rangle+\mathcal O(\Psi^4)$, subject to $B_1^2=0$ and $[B_1,B_2]=0$ on fields, gauge parameters and Noether identities. The gravitational vertices obtained here are representatives of interactions which admit consistent $L_\infty$ completions, organized as $B_1=B_1^{(0)}+\alpha'B_1^{(1)}$ and $B_2=B_2^{(0)}+\alpha'B_2^{(1)}+\alpha'^2B_2^{(2)}$. For instance, the curvature-cubic terms do not modify $B_1$, and since at cubic order they are built from linearized curvatures, they define a separately gauge-invariant two-product $B_2^{(2)\mu\nu}(h_1,h_2)=4\partial_\rho\partial_\sigma P^{\mu\rho\sigma\nu}(h_1,h_2)$, where $P$ is the symmetric polarization of the derivative of the Riem$^3$ invariants with respect to the curvature, evaluated on linearized curvatures. At order $\alpha'$, instead, the curvature-squared term corrects both the linearized equations, $B_1^{(1)\mu\nu}(h)=\partial_\rho\partial_\sigma R^{(1)\mu\rho\sigma\nu}(h)$, and the cubic vertices, and the two corrections are related by
\be
[B_1^{(0)},B_2^{(1)}]+[B_1^{(1)},B_2^{(0)}]=0\, ,
\label{LinftyOrderOne}
\ee
which explains why the quadratic Riem$^2$ term, although removable from the comparison in TT gauge, must be retained in an unredefined off-shell algebra. It would be interesting to extend this analysis from the Riem$^2$ term to the full four-derivative bosonic supergravity, including the $\mathrm{Riem}\,H^2$ coupling reproduced in section~\ref{Sec:B}, the terms quartic in $H$, and the dilaton couplings. In the scheme \eqref{bosonicintro} the gauge transformations are not deformed, while in duality-covariant formulations of the same theory, such as the generalized Bergshoeff--de Roo identification \cite{Bergshoeff:1989de,Marques:2015vua}, the transformations of the Kalb--Ramond field acquire $\alpha'$ corrections, and the two descriptions are related by field redefinitions. In the $L_\infty$ language these formulations should correspond to quasi-isomorphic algebras, and it would be interesting to determine which of them is selected by the deformed replacement rules, given that the double copy is naturally formulated in terms of the doubled field $e_{\mu\bar\nu}$. Such a construction requires the off-shell completion of the momentum tensor $\Delta\bar\pi^{(3)}$ and the quadratic corrections induced by the $(DF)^2$ operator, and it would establish whether the deformed replacement rule preserves the off-shell gauge invariance of the construction of \cite{Diaz-Jaramillo:2021wtl}.

Finally, the cubic analysis can be completed with the dilaton, which is set to zero by the gauge conditions used here, and extended to the heterotic string. The asymmetric solutions of section~\ref{Sec:alpha2} reproduce the pure gravitational pattern of the heterotic string, and the $hhb$ coupling they generate is the one associated with the Lorentz Chern--Simons term. A complete treatment requires the inclusion of the gauge sector of the heterotic string, and would provide a further connection with the generalized Bergshoeff--de Roo deformations of duality-covariant formulations \cite{Bergshoeff:1989de,Marques:2015vua}.

\acknowledgments
The work of E.L. is supported by the SQUASH program, co-funded by the European Union's Marie Sk\l{}odowska-Curie COFUND action (Grant Agreement PR-14104-2) and the Republic of Slovenia. A.P.G. and J.A.R. acknowledge support from CONICET. J.A.R. also acknowledges support from Universidad Argentina de la Empresa (UADE) through the Academic Research Project ``Black Hole Physics in Modern Theories of Gravity'' (A26T65). The results of this project were checked with GPT-6 (OpenAI) and Fable 5.1 (Anthropic). The construction of the double copy action was done with Cadabra \cite{Peeters:2006kp,Peeters:2007wn}. The code is available upon request.

\appendix

\section{Cubic vertices}
\label{App:support}

In this appendix we describe the method used to perform the integrations by parts and field redefinitions in the main text, and collect the reductions of the relevant cubic structures.

\paragraph{Method.} Let $L^{(3)}$ be a local cubic Lagrangian in the fields $e_{\mu\bar\nu}$, or $A_\mu{}^a$ on the gauge-theory side. We insert
\be
e_{\mu\bar\nu}(x) = \sum_{i=1}^{3} t_i\, e_{i\,\mu\bar\nu}\, e^{ik_i\cdot x}\, ,
\qquad
A_\mu{}^a(x)=\sum_{i=1}^3 t_i\,\varepsilon_{i\mu}\,T_i^a\,e^{ik_i\cdot x}\, ,
\ee
with $t_i^2=0$, and extract the coefficient of $t_1t_2t_3$, which we denote by $\mathcal{E}[L^{(3)}]$. Since $k_1+k_2+k_3=0$, this coefficient is independent of $x$, so that total derivatives do not contribute. We impose
\be
k_i\cdot k_j=0\, ,\qquad k_i^{\mu}e_{i\,\mu\bar\nu}=0=e_{i\,\mu\bar\nu}k_i^{\bar\nu}\, ,\qquad \eta^{\mu\bar\nu}e_{i\,\mu\bar\nu}\big|_{h}=0\, ,\qquad \varepsilon_i\cdot k_i=0\, ,
\ee
where the trace condition applies to the symmetric part. For three real Lorentzian momenta these conditions force the $k_i$ to be collinear and all vertices to vanish; as in the standard treatment of three-point amplitudes, we therefore take the momenta and polarizations to be complex. The reductions obtained in this way are polynomial identities, and hold for arbitrary real fields once translated back to position space. The conditions $k_i\cdot k_j=0$ remove precisely the terms proportional to $\Box$ acting on a single field, and the remaining ones implement the gauge \eqref{TT}. Two cubic Lagrangians are equal up to integrations by parts and field redefinitions in this gauge if and only if their values $\mathcal{E}$ coincide, and we denote this by $\cong$. On the gauge-theory side we take $T_i^a$ such that $f_{abc}T_1^aT_2^bT_3^c=1$.

\paragraph{Building blocks.} On the support, every cubic vertex is a polynomial in the tensors
\be
T_{\mathrm{YM}}^{\mu\nu\rho} = \eta^{\mu\nu}k_1^{\rho}+\eta^{\nu\rho}k_2^{\mu}+\eta^{\rho\mu}k_3^{\nu}\, ,\qquad
T_{F^3}^{\mu\nu\rho} = k_2^{\mu}k_3^{\nu}k_1^{\rho}\, ,
\label{Ttensors}
\ee
and their barred counterparts. For factorized polarizations $e_{i\,\mu\bar\nu}=\varepsilon_{i\mu}\bar\varepsilon_{i\bar\nu}$ we define
\be
\cA_{\mathrm{YM}} = T_{\mathrm{YM}}^{\mu\nu\rho}\varepsilon_{1\mu}\varepsilon_{2\nu}\varepsilon_{3\rho}\, ,\qquad
\cA_{F^3} = T_{F^3}^{\mu\nu\rho}\varepsilon_{1\mu}\varepsilon_{2\nu}\varepsilon_{3\rho}\, ,
\ee
and similarly $\bar\cA_{\mathrm{YM}}$, $\bar\cA_{F^3}$; the pure gravitational sector corresponds to $\bar\varepsilon_i=\varepsilon_i$ with $\varepsilon_i^2=0$. The general case is recovered by multilinearity, replacing products such as $\cA\bar\cA$ by the full contraction $T^{\mu\nu\rho}\bar T^{\bar\mu\bar\nu\bar\rho}e_{1\mu\bar\mu}e_{2\nu\bar\nu}e_{3\rho\bar\rho}$. On the support one finds
\be
\pi^{(3)\mu\nu\rho}\cong 2T_{\mathrm{YM}}^{\mu\nu\rho}\, ,\qquad \Delta\pi^{(3)\mu\nu\rho}\cong 2T_{F^3}^{\mu\nu\rho}\, .
\ee

\paragraph{Reductions.} The cubic structures used in the main text reduce as follows:
\begin{center}
\renewcommand{\arraystretch}{1.3}
\begin{tabular}{ll}
\toprule
Cubic Lagrangian & $\mathcal{E}$ on the support \\
\midrule
$\big(\sqrt{-g}R\big)^{(3)}$, \ $L^{(3)}_{\mathrm{DC}}|_h$ in \eqref{DC_cubic_h} & $\frac{1}{2}\cA_{\mathrm{YM}}^2$ \\
$\big(\sqrt{-g}\,R_{\mu\nu\rho\lambda}R^{\mu\nu\rho\lambda}\big)^{(3)}$, \ $\mathcal{S}_1$ & $2\,\cA_{\mathrm{YM}}\cA_{F^3}$ \\
$\partial_\mu\partial_\nu h^{\rho}{}_{\lambda}\partial_\sigma\partial^\mu h^{\lambda}{}_{\rho}h^{\nu\sigma}$ & $0$ \\
$R^{\mu}{}_{\nu}{}^{\rho\lambda}R^{\nu}{}_{\sigma\lambda}{}^{\tau}R^{\sigma}{}_{\rho\tau\mu}$ & $\frac{3}{2}\cA_{F^3}^2$ \\
$R^{\mu\nu}{}_{\rho\lambda}R^{\sigma}{}_{\tau}{}^{\rho\lambda}R^{\tau}{}_{\mu\sigma\nu}$ & $-3\,\cA_{F^3}^2$ \\
$\mathcal{O}_1$ & $-6\,\cA_{F^3}^2$ \\
$L^{(3)}_{\mathrm{NS}}$, \ $L^{(3)}_{\mathrm{DC}}$ in \eqref{DC_cubic} & $\frac{1}{2}\,T_{\mathrm{YM}}\bar T_{\mathrm{YM}}$ \\
$R^{(1)}_{\beta\gamma\delta\epsilon}(h)\,H_{\alpha}{}^{\delta\epsilon}H^{\alpha\beta\gamma}$ & $-2\,\big(T_{\mathrm{YM}}\bar T_{F^3}+T_{F^3}\bar T_{\mathrm{YM}}\big)$ \\
\midrule
$-\frac{1}{4}\kappa_{ab}F_{\mu\nu}{}^aF^{\mu\nu b}$ & $i\,\cA_{\mathrm{YM}}$ \\
$\kappa_{ab}D^{\nu}F^{\sigma\epsilon a}D_{\sigma}F_{\nu\epsilon}{}^{b}$ & $6i\,\cA_{F^3}$ \\
$f_{abc}F_{\mu\nu}{}^{a}F^{\nu\rho b}F_{\rho}{}^{\mu c}$ & $-12i\,\cA_{F^3}$ \\
$D^nFD^nF$, \ $n\geq 2$ (e.g.~\eqref{Newbi}) & $0$ \\
\bottomrule
\end{tabular}
\end{center}
where the gauge-theory entries are evaluated with $g_{\mathrm{YM}}=\frac{1}{2}$, and in the last two gravitational rows $T\bar T$ stands for the full contraction $T^{\mu\nu\rho}\bar T^{\bar\mu\bar\nu\bar\rho}e_{1\mu\bar\mu}e_{2\nu\bar\nu}e_{3\rho\bar\rho}$ with general transverse polarizations $e_{i\,\mu\bar\nu}=h_{i\,\mu\nu}+b_{i\,\mu\nu}$. The two-derivative row is the first term of \eqref{Bcubic} together with its pure-gravity part, and the $R\,H^2$ row shows that the $-\frac18 R^{(1)}HH$ term of \eqref{Bcubic} equals $\frac14(T_{\mathrm{YM}}\bar T_{F^3}+T_{F^3}\bar T_{\mathrm{YM}})$, which is precisely the order-$\alpha'$ part of \eqref{symmetric}. For $hhb$ polarizations $T_{\mathrm{YM}}\bar T_{F^3}=-T_{F^3}\bar T_{\mathrm{YM}}\neq0$, which is the origin of the $hhb$ coupling of the asymmetric solutions discussed in section~\ref{Sec:B}. The two gauge-theory entries for $(DF)^2$ and $F^3$ are consistent with the identity \eqref{DFidentity}, and show that the cubic gauge-theory vertex is proportional to $\cA_{\mathrm{YM}}-12\lambda\,\cA_{F^3}$, in agreement with \eqref{gaugeVertex}. With the normalization of \eqref{cubic}, the double-copy action \eqref{DCdeformed} gives
\be
\mathcal{E}\big[L^{(3)}_{\mathrm{DC}}\big] = \frac{1}{2}\left(\bar\cA_{\mathrm{YM}}+4c_1\bar\cA_{F^3}\right)\left(\cA_{\mathrm{YM}}-12\lambda\,\cA_{F^3}\right)\, ,
\ee
from which \eqref{DeltaContributionGeneral} follows using the table. For the values \eqref{solution}, this is $\frac{1}{2}(\bar\cA_{\mathrm{YM}}+\frac12\bar\cA_{F^3})(\cA_{\mathrm{YM}}+\frac12\cA_{F^3})$. The same method, applied to arbitrary symmetric and antisymmetric transverse polarizations, gives the results of section~\ref{Sec:B}.

\bibliographystyle{JHEP}
\bibliography{references}

\end{document}